\documentclass[useAMS,usegraphicx]{mn2e}
\usepackage{epsfig}
\newcommand{\asec}{$^{\prime\prime}$}
\newcommand{\pas}{.\hskip-2pt$^{\prime\prime}$}
\def\ii{I19520}
\newcommand{\Td}    {T_\mathrm{d}}
\newcommand{\Tex}   {T_\mathrm{ex}}

\newcommand{\Tbg}    {T_\mathrm{bg}}
\newcommand{\mum}   {$\mu$m}
\newcommand{\kms}   {km~s$^{-1}$}

\newcommand{\cmt}   {cm$^{-3}$}
\newcommand{\jpb}   {$\rm Jy~beam^{-1}$}    
\newcommand{\lo}    {$L_{\sun}$}
\newcommand{\mo}    {$M_{\sun}$}
\newcommand{\Av}    {$A_\mathrm{v}$}
\newcommand{\co}    {$^{12}$CO}
\newcommand{\tco}    {$^{13}$CO}
\newcommand{\ceo}    {C$^{18}$O}

\newcommand{\water}  {H$_2$O}

\newcommand{\et}    {et al.}
\newcommand{\eg}    {e.\,g.,}
\newcommand{\ie}    {i.\,e.,}

\newcommand{\supa}  {$^\mathrm{a}$}
\newcommand{\supb}  {$^\mathrm{b}$}
\newcommand{\supc}  {$^\mathrm{c}$}

\newcommand{\phb}   {\phantom{$>$}}

\title[Millimetre emission from IRAS 19520+2759]
{IRAS 19520+2759: a 10$^5$~\lo\ massive young stellar object driving a collimated outflow}

\author[Palau et al.]{
Aina Palau$^{1}$\thanks{E-mail: palau@ieec.uab.es}, 
C. S\'anchez Contreras$^{2}$, R. Sahai$^3$, \'A. S\'anchez-Monge$^{4}$, J. R. Rizzo$^{2}$
\\
$^{1}$ Institut de Ci\`encies de l'Espai (CSIC-IEEC), Campus UAB Facultat de Ci\`encies,Torre C5-parell 2, E-08193 Bellaterra, Catalunya, Spain\\
$^{2}$ Centro de Astrobiolog\'{\i}a (INTA-CSIC), Ctra. M-108, km. 4, E-28850 Torrej\'on de Ardoz, Madrid, Spain\\
$^{3}$ Jet Propulsion Laboratory, MS 183-900, California Institute of Technology, Pasadena, CA 91109\\
$^{4}$ INAF-Osservatorio Astrofisico di Arcetri, Largo E. Fermi 5, I-50125, Firenze, Italy\\
}
\begin{document}

\date{Accepted date. Received date; in original form date}

\pagerange{\pageref{firstpage}--\pageref{lastpage}} \pubyear{2012}

\maketitle

\label{firstpage}

\begin{abstract}
The theory of massive star formation currently suffers from a scarce observational base of massive young stellar objects to compare with. In this paper, we present interferometric \co\,(1--0), \tco\,(1--0), \ceo\,(1--0), and 2.6~mm continuum images of the infrared source IRAS\,19520+2759 together with complementary single-dish observations of CS\,(1--0), obtained with the 34\,m antenna DSS-54 at the Madrid Deep Space Communications Complex, as well as archive images at different wavelengths. As a result from our work, IRAS\,19520+2759, with a controversial nature in the past, is firmly established as a massive young stellar object associated with a strong and compact millimetre source and driving a collimated outflow. In addition, a second fainter millimetre source is discovered about 4~arcsec to the south, which is also driving an outflow. Furthermore, the two millimetre sources are associated with \ceo\ clumps elongated perpendicularly to the outflows, which may be related to rotating toroids. The masses of gas and dust of the millimetre sources are estimated to be around 100 and 50~\mo. MM1, the dominant source at all wavelengths, with a total luminosity of (1--2)$\times10^5$~\lo\ at 9~kpc, is however not associated with 6~cm emission down to a rms noise level of 0.1 mJy. We propose that IRAS\,19520+2759 could be an example of the recent theoretical prediction of `bloated' or `swollen' star, \ie\ a massive young stellar object whose radius has increased due to effects of accretion at a high-mass accretion rate.
\end{abstract}


\section{Introduction}\label{intro}

\begin{figure*}
\begin{center}
\begin{tabular}[b]{c}
     \epsfig{file=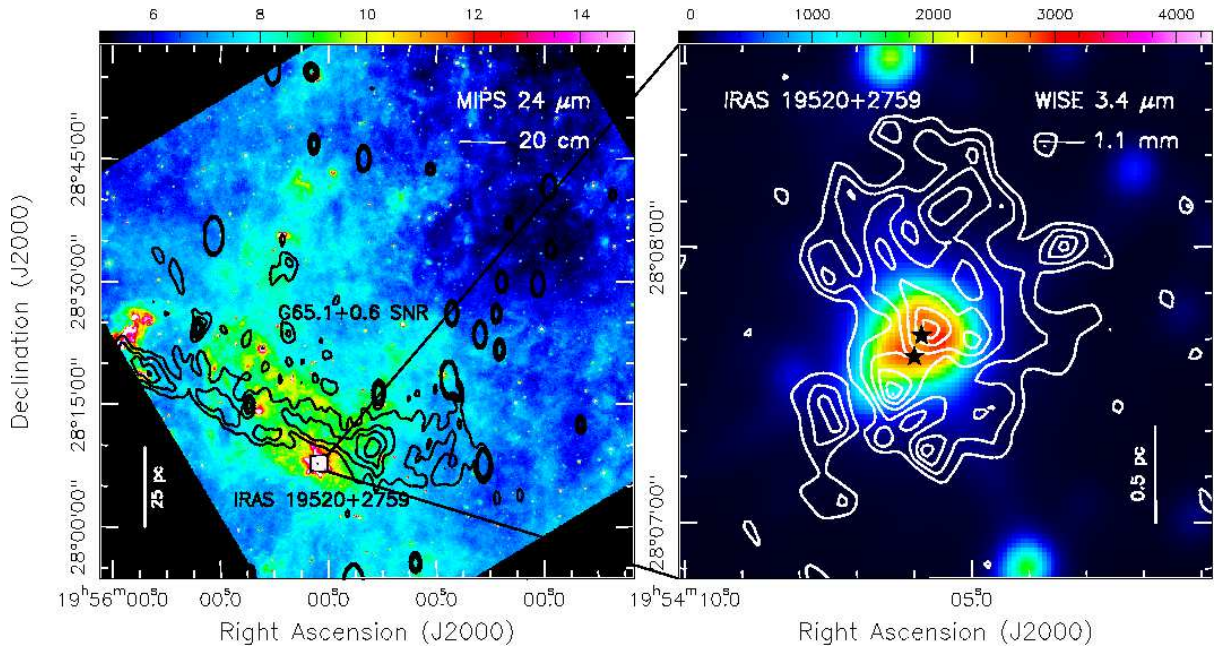,  width=16cm, angle=0} \\
\end{tabular}
\caption{{\bf Left:} Colorscale: Spitzer/MIPS at 24~\mum\ showing the surroundings of \ii\ (colorscale units in MJy\,sr$^{-1}$). Black contours: 20~cm emission from Landecker \et\ (1990) and Kothes \et\ (2006) tracing the Supernova Remnant (SNR). Contours are 7.5, 7.7, 7.9, and 8.1~K. Note that the 24~\mum\ emission is extended at the expanding front of the SNR and that \ii, located about 4 arcmin to the south, is the most prominent source in the field. {\bf Right:} Zoom-in on the \ii\ region, with the colorscale (in arbitrary units) tracing emission at 3.4~\mum\ from the Wide-field Infrared Survey Explorer (WISE, Wright \et\ 2010) and the white contours (of 0.12, 0.16, 0.20, 0.24, 0.28, and 0.32~\jpb) corresponding to the 1.1~mm Bolocam emission (Rosolowsky \et\ 2010; Aguirre \et\ 2011). The two black stars indicate the position of the two millimetre sources detected in this work (Section~\ref{srcont}). The linear scale, indicated by the vertical segment within the panels, is computed adopting a distance of 9~kpc to \ii\ (Section~1).
}   
\label{flargefov}
\end{center}
\end{figure*}

Although high-mass ($>8$~\mo) stars are clue pieces in the Universe and galaxy evolution, many questions remain open concerning their formation process.  The root of the problem is that the Kelvin-Helmholtz timescale for a high-mass star is much shorter than the free-fall timescale of the natal core, and thus the star reaches the main-sequence while still accreting matter. The important amounts of UV radiation emitted by the nascent massive star should halt at some point further accretion onto the star, limiting its final mass to a certain value which is currently not well established. From a theoretical point of view, there are a number of studies showing that it is possible to form a 40--50~\mo\ star through disc-mediated accretion (\eg\ Yorke \& Sonnhalter 2002; Krumholz \et\ 2009),  and even it seems possible to form stars of up to 140~\mo\ (\eg\ Kuiper \et\ 2010, 2011).
However, from an observational point of view, a clear picture of the different evolutionary stages that a massive star undergoes until reaching the main-sequence is lacking, mainly due to the fast evolution of massive stars. 
Beuther \& Shepherd (2005) and Keto (2007) suggest an observational sequence for massive star formation, where the O-type stars progenitors are B-type objects embedded in large amounts of dust and gas and accreting at large mass infall rates (10$^{-4}$--10$^{-3}$~\mo\,yr$^{-1}$). In this scenario, the B-type object drives a collimated molecular outflow, and while progressively moving from early-B to late-O and finally early-O, the opening angle of the outflow progressively increases.
Thus, the earliest stages in the formation of an O-type star should be characterized by a central B-type object, driving a collimated outflow, and embedded in an infalling envelope massive enough to provide the B-type object with the required mass to become an O-type object. Typically, the massive young stellar objects (YSOs) in the earliest evolutionary stages also present emission from complex organic molecules (the so-called hot molecular cores, \eg\ Cesaroni \et\ 2010) and a lack of strong centimetre emission (\eg\ Beuther \et\ 2002; Molinari \et\ 1998; Mottram \et\ 2011). A possible reason for the lack of an HII region is the quenching and/or trapping of the ionized gas by the infalling dense envelope (\eg\ Walmsley 1995; Molinari \et\ 1998; Keto 2002, 2003), or, as recently proposed by  Hoare \& Franco (2007), an increase of stellar radius as accretion proceeds onto the star due to trapped entropy (the so-called `swollen' or `bloated' stars), which produces a decrease of the stellar effective temperature and of the UV radiation emitted by the nascent star (\eg\ Yorke \& Bodenheimer 2008; Hosokawa, Yorke, \& Omukai 2010). 


Thus, the study of outflow and accretion signatures in massive YSOs is essential to progress in the understanding of massive star formation.
However, there are only a few O-type
%
YSO candidates (\ie\ with bolometric luminosities $\ga50000$~\lo) associated with disc-like structures or toroids of dense gas and possibly undergoing infall (\eg\ Zhang, Ho, \& Ohashi 1998; Pestalozzi \et\ 2004; Sollins \& Ho 2005, Sollins \et\ 2005a; Beltr\'an \et\ 2006; Beuther \et\ 2007a,b; Beuther \& Walsh 2008; Franco-Hern\'andez \et\ 2009; Cesaroni \et\ 2011; Furuya \et\ 2011; Qiu \et\ 2012; Jim\'enez-Serra \et\ 2012; Wang \et\ 2012),
and it is not clear if these massive YSOs harbor true single O-type objects or a multiple system of early-B objects (\eg\ Naranjo-Romero \et\ 2012). This cannot be easily elucidated because most of the aforementioned massive YSOs undergoing infall are not detected at optical wavelengths and their spectral type cannot be assessed.
On the other hand, YSOs with luminosities $\ga50000$~\lo\ and detected in the optical/NIR have been found associated with accreting disks through near-infrared observations (\eg\ Jiang \et\ 2008; Davies \et\ 2010; de Wit \et\ 2010; Kraus \et\ 2010; Wheelwright \et\ 2012), but their number is also scarce.
Thus it is extremely necessary to enlarge the sample of O-type YSOs undergoing infall and actively accreting, and simultaneously detected in the optical/NIR to characterize not only the infalling envelope but also the central object. The results obtained from the joint analysis of multiwavelength data presented in this paper indicate that one new actively accreting O-type (luminosity $\sim10^5$~\lo) candidate is IRAS\,19520+2759 (R.A.\,(J2000)=19$^{\rm h}$54$^{\rm m}$05\fs9 and Dec.\,(J2000)=$+$28\degr07\arcmin41$''$).

The status of IRAS\,19520+2759 (hereafter, \ii) has been controversial: it was
originally suggested to be an OH/IR star based on its infrared colors
(Habing \& Olnon 1984). However, Arquilla \& Kwok (1987) examined the InfraRed Astronomical Satellite 
(IRAS) images towards this region and found the far-infrared emission
to be extended. This, together with its spectral energy distribution
(nearly flat from NIR to 100~$\mu$m), led these authors to conclude
that this object was likely a young star embedded in an extended
molecular cloud.  In addition, Arquilla \& Kwok (1987) present a
single-dish \co\ spectrum with broad wing emission (FWZI~$\sim$~30~\kms),
similar to the line wings associated with outflows in star-forming regions, 
and Gledhill (2005) reports the detection of a polarized reflection nebula in the near-infrared.
\water\ and OH maser emission with single-peaked profiles
(\ie\ unlike double-peaked profiles in evolved stars) lends
additional credence to the identification of this source as a YSO (Engels et al.\,1984, Lewis et al.\ 1985). Moreover,
the non detection of SiO maser emission (Nakashima \& Deguchi 2003) is
also consistent with the YSO nature (\eg\ Zapata \et\ 2009b). 

\ii\ stands out in the infrared as the brightest and reddest object
within a field of view of $\sim50$~arcmin. A recent image from the Spitzer MIPSGAL survey (Multiband Imaging Photometer for Spitzer Inner Galactic Plane Survey, Carey \et\ 2009) at 24~\mum\ shows a very
bright, compact source coincident with the IRAS source and
surrounded by a much more extended ($\ge$$15'$), diffuse emission tracing
a large molecular cloud (see Fig.~\ref{flargefov}). Cold dust emission from this large
molecular cloud is traced by Bolocam observations at 1.1\,mm (Rosolowsky \et\ 2010; Aguirre \et\ 2011), which
reveal an extended ($\sim40$~arcsec) source, of about 0.7~Jy of flux
density, where \ii\ is embedded (Fig.~\ref{flargefov}). Interestingly,
\ii\ is located at the boundary of the supernova remnant G\,65.1+0.6 in a
region where the remnant is probably interacting with the surrounding
interstellar medium (Fig.~\ref{flargefov}; Landecker et al.\,1990; Tian \& Leathy 2006).

Hrivnak et al. (1985) identified an optical counterpart candidate of
\ii, which appears as a bright, point-like source as observed with the Hubble Space Telescope (GO 9463, PI: R.\,Sahai).
Recently, S\'anchez Contreras \et\ (2008) report echelle long-slit 3900--10900~\AA\ spectroscopy of \ii\footnote{Due to its similar IRAS colors, \ii\ was included in the list of pre Planetary Nebula candidates by these authors.},
and show that the optical spectrum is dominated by emission lines consistent with the presence of a compact, photoionized region around an early-type (B or ealier) star (an in depth study of the optical spectrum is in preparation by S\'anchez Contreras et al.).
Finally, the intense H$\alpha$ emission detected toward the nebular core shows broad wings ($\sim$2600~\kms) and a remarkable P-Cygni profile with \emph{two} blue-shifted absorption features, suggesting multiple stellar winds with (projected) expansion velocities of up to 800~\kms. 

As already noted by Arquilla \& Kwok (1987), the galactic longitude
and negative V$_{\rm LSR}$ of \ii\ indicate that this object is very
remote. We estimate a value for its kinematic distance of $d_{\rm k}\sim8.8$\,kpc, determined from the target radial velocity (V$_{\rm
  LSR}$$\sim$$-$16.5\,km/s) and its galactic coordinates
($l$=64.8131\degr, $b$=+00.1744\degr) by assuming a simple galactic
rotation law and adopting a value for the A Oort constant of
14.4\,\kms\,kpc$^{-1}$ and a galactocentric radius of 8.5\,kpc (Kerr
\& Lynden-Bell, 1986). As indicated by Arquilla \& Kwok (1987), this
value of the distance is consistent with the non-kinematic distances,
8--10\,kpc, of two H\,II regions in the catalog of Blitz et
al.\, (1982) that are likely to occupy approximately the same region
of the Galaxy (given their similar $l$ and radial velocities). The range of
radial velocities displayed by the H\,I 21\,cm-line emission profile, V$_{\rm LSR}$=[$-$20:$-$26\,km/s], 
at the boundary of the SNR G65.1+0.6, next to which \ii\ and the
surrounding molecular cloud are located (Fig. 1), consistently yield a
kinematic distance for the SNR of 9--10\,kpc (Landecker et al.\,1990; Tian \& Leahy 2006).
In view of these results, we adopt a distance of 9~kpc for \ii. 




The rest of this paper is organized as follows. In Section~2, we describe the observations, in Section~3 we give the main results from the OVRO 2.6~mm, and \co(1--0), \tco(1--0) and \ceo(1--0), in Section~4 we derive the main physical properties of the molecular gas, outflow and central object in \ii, and in Section~5 we discuss a possible interpretation for \ii\ within the massive star formation framework.

\section{Observations and data reduction}\label{obs}

\subsection{OVRO observations}

Interferometric mapping of the $^{12}$CO, $^{13}$CO, and C$^{18}$O
$J$=1$-$0 line emission at 108--115\,GHz of \ii\, was carried out using the six 10.4-m
antennas millimetre array of the Owens Valley Radio Observatory
(OVRO), which is now part of The Combined Array for Research in
Millimeter-wave Astronomy (CARMA\footnote{\tt
  http://www.mmarray.org}). Observations were performed 
in different runs between 2002 and 2004 using the compact (C),
equatorial (E), and high (H) array configurations, with baselines
ranging between 4.01 and 93.3 k$\lambda$ in the $uv$ plane. The
coordinates of the tracking center are R.A.\,(J2000)=19$^{\rm
  h}$54$^{\rm m}$05\fs92 and Dec.\,(J2000)=$+$28\degr07\arcmin41\farcs5, and the total on-source time was $\sim$18\,hr.

\begin{table}
\small 
\caption{Parameters of line and 2.6\,mm continuum OVRO interferometric maps}
\begin{center}
\begin{tabular}{lcccc}
\hline
\hline 
& Array  
& Beam size  
&  P.A. 
& rms\supa /rms\supb  
\\
Maps
& Config.
&(\arcsec$\times$\arcsec)  
&(\degr) 
&(mJy\,beam$^{-1}$) 
\\
\hline 
$^{12}$CO\,(1--0) 		&CEH	&$4.91\times4.34$	&85.5	&17/63\\
$^{13}$CO\,(1--0) 		&CEH	&$5.36\times4.57$	&72.7	&12/25\\
C$^{18}$O\,(1--0)	 	&CEH	&$5.56\times4.54$	&76.7	&10/12\\
2.6\,mm cont.			&CEH	&$4.53\times3.60$	&83.6	&0.5\\
2.6\,mm cont. 			&H		&$2.29\times1.45$	&81.8	&0.7\\
\hline	     
\end{tabular}
\begin{list}{}{}
\item[$^\mathrm{a}$] As measured in 1\,MHz-wide channels with no signal 
\item[$^\mathrm{b}$] As measured in the central 1\,MHz-wide channel where the CO brightness distribution peaks
\item[]
\end{list}
\end{center}
\label{tobs}
\end{table}

\begin{table*}
\caption{Parameters of the sources detected with OVRO at 2.6~mm using the H configuration (highest angular resolution)}
\begin{center}
{\small
\begin{tabular}{lccccccc}
\noalign{\smallskip}
\hline\noalign{\smallskip}
&\multicolumn{2}{c}{Position$^\mathrm{a}$}
&Dec. ang. size
&Dec. P.A.
&$I_\mathrm{\nu}^\mathrm{peak}$~$^\mathrm{a}$
&$S_\mathrm{\nu}$~$^\mathrm{a}$
&Mass$^\mathrm{b}$
\\
\cline{2-3}
Source
&$\alpha (\rm J2000)$
&$\delta (\rm J2000)$
&$(''\times'')$
&($^\circ$)
&(m\jpb)
&(mJy)
&(\mo)\\
\noalign{\smallskip}
\hline\noalign{\smallskip}
MM1       	&19:54:05.87   &28:07:40.9 &\phb$2.4\times0.7$ &$-72$		&$9.0\pm0.7$	&$16\pm5$      	&110\\
MM2       	&19:54:06.00   &28:07:36.3 &$<1.1\times0.7$	&-    			&$4.8\pm0.7$	&$7\pm2$		&48\\
MM2	W	&19:54:05.82   &28:07:36.3 &$<1.1\times0.7$	&-	    		&$3.4\pm0.7$	&$3\pm1$       	&20\\
\hline
All$^\mathrm{c}$&19:54:05.87 &28:07:41.1 &$3.4\times 2.8$   &52		&$13.0\pm0.5$ &$20\pm6$	&140\\
\hline
\end{tabular}
\begin{list}{}{}
\item[$^\mathrm{a}$] Position, peak intensity and flux density are derived by fitting a Gaussian in the image domain (double gaussian for the case of MM2 and MM2W).
Uncertainty in the peak intensity is the rms noise of the cleaned image, $\sigma$.  
Error in flux density has been calculated as $\sqrt{(\sigma\,\theta_\mathrm{source}/\theta_\mathrm{beam})^{2}+(\sigma_\mathrm{flux-scale})^{2})}$ (Beltr\'an \et\ 2001), where $\theta_\mathrm{source}$ and $\theta_\mathrm{beam}$ are the size of the source and the beam respectively, and $\sigma_\mathrm{flux-scale}$ is the error in the flux scale, which takes into account the uncertainty on the calibration applied to the flux density of the source ($S_\nu\times\%_\mathrm{uncertainty}$).
\item[$^\mathrm{b}$] Masses derived assuming a dust temperature of 50~K (from a grey-body fit to the SED at wavelengths $>100$~\mum), and a dust (and gas) mass opacity coefficient of 0.003~cm$^2$\,g$^{-1}$ (obtained by extrapolating the tabulated values of Ossenkopf \& Henning 1994, see Section~\ref{srcont}). The uncertainty in the masses due to the dust temperature and opacity law is estimated to be a factor of 4.
\item[$^\mathrm{c}$] Parameters derived from the image built using C+E+H configurations. For this case the peak intensity and flux density are derived by computing statistics in a region of 12~arcsec of diameter.
\end{list}
}
\end{center}
\label{tcont}
\end{table*}

Observations of the $^{12}$CO, $^{13}$CO, and C$^{18}$O ($J$=1--0)
transitions were performed simultaneously.  The digital spectral line
correlator was configured to provide a total bandwidth of 3$\times$32\,MHz
($\sim$\,250\,\kms) with a channel spacing of 1\,MHz
($\sim$\,2.6\,\kms) in the $^{12}$CO and $^{13}$CO lines. For
C$^{18}$O, the units of the cross-correlator were set to bandwidths of
32\,MHz ($\sim$85\,\kms) with channel spacing of 1\,MHz (2.7\,\kms).
The 2.6\,mm continuum emission was observed simultaneously using the
dual-channel analog continuum correlator. Although the continuum
correlator provides a total bandwidth of 4\,GHz after combining both
IF bands, our continuum maps have a bandwidth of 3\,GHz since one of
the 1\,GHz bands, which contained the $^{12}$CO emission line, has not
been used to generate the final maps. 

Data calibration was performed using the MMA software package (Scoville \et\ 1993). Gain-calibration was done against a nearby quasar,
J2025+337, which was observed at regular time intervals of
20\,minutes before and after each target observation. The
bright quasar 3C\,454.3 was observed at the beginning or end of the track
for passband calibration. Absolute flux calibration was obtained by
observing Uranus and 3C\,454.3 as primary and secondary flux
calibrators, respectively. Flux calibration errors could be of up to 20--30\%, and
absolute positional accuracies are estimated to be $\sim0.2$~arcsec.

Reconstruction of the maps from the visibilities was done using
standard tasks of the Multichannel Image Reconstruction, Image
Analysis and Display (MIRIAD) software (Sault \et\ 1995). After Fourier transforming the
measured visibilities with natural weighting, data were cleaned and
maps restored. 
For the continuum, we have separately obtained two sets of maps, one
set using all the data available from all three, C+E+H, configurations,
and another one using only data from the H configuration; the later 
provides maps with improved spatial resolution but lower S/N. For the line emission, imaging of the H configuration data suffers from an important decrease of S/N and missing flux, and for this reason we only show here the C+E+H configuration data for the line. 
The size and orientation of the {\bf synthesized} clean beam together with the noise 
in our maps are given in Table~\ref{tobs}.



\subsection{Single-dish observations with DSS-54 of Madrid DSCC}

We have used the DSS-54 antenna of NASA's Madrid Deep Space Communications 
Complex (DSCC) in Robledo de Chavela, under the `Host Country' program. 
The antenna is a beam-wave-guide paraboloid, with a Coud\'e focus, and has 
a diameter of 34\,m. The observations were done using the dual-polarization 
Q-band High Electron Mobility Transistor (HEMT) receiver attached to the antenna. We have observed both circular 
polarizations using the new wideband backend (Rizzo \et\ 2012a).

The CS\,(1--0) line (48.990955\,GHz) was observed using two spectral windows having bandwidths of 500\,MHz 
and 16,384 channels, which provide a spectral resolution of 30.5\,kHz 
(0.19\,km\,s$^{-1}$ at the line frequency). Both 500-MHz bands were centred 
at 48.82\,GHz in order to profit the wideband capabilities of the backend, and search for other molecular 
lines. Pointing was regularly checked, and was found to be better than 6~arcsec.
The Half Power Beam Width (HPBW) of the antenna at the observed frequency is 44~arcsec.
Observations were done in position-switching mode, with a reference located 3.6~arcmin in azimuth.

Data were gathered on April 25, 2012, under good weather conditions; 
atmospheric opacity at the observed frequency was close to 0.1. The total on-source
integration time was 46~minutes, and the system temperature was 
less than 180\,K. Estimated {\it rms} noise level was 40\,mK, on a main-beam 
temperature scale. Control of the wideband backend and synchronization of the whole observing processes were done using the SDAI software (Rizzo \et\ 2012b). The gathered data were written into FITS files by SDAI, and were subsequently read and further 
processed using the CLASS software\footnote{CLASS is part of the GILDAS software, developed by the Institute de Radioastronomie Millimetrique (IRAM). See
http://www.iram.fr/IRAMFR/GILDAS.}.

\begin{figure}
\begin{center}
\begin{tabular}[b]{c}
     \epsfig{file=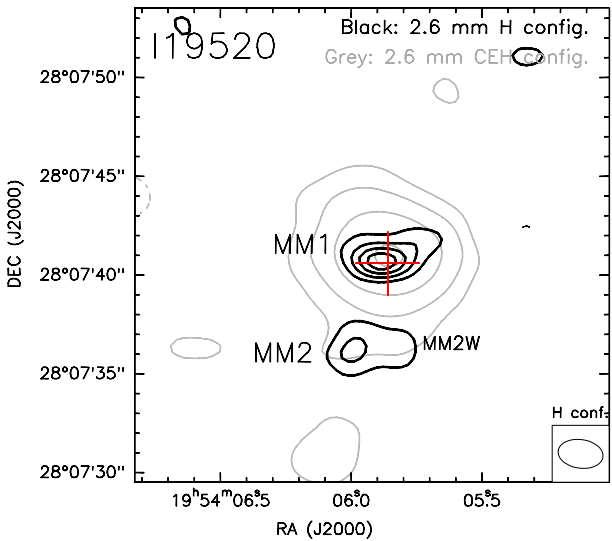,  width=8cm, angle=0} \\
\end{tabular}
\caption{
Black contours: OVRO 2.6~mm continuum emission observed with the H configuration. Black contours are $-3$, 3, 6, 9, and 12 times the rms of the map, 0.7~m\jpb. The synthesized beam of the H configuration data is shown in the bottom-right corner ($2.29\times1.45$~arcsec$^2$, P.A.= 81.8\degr).
Grey contours: OVRO 2.6~mm continuum emission obtained by combining three configurations (C+E+H), providing a synthesized beam of $4.53\times3.60$~arcsec$^2$, P.A.=83.6\degr. Contours are $-3$, 3, 6, 12, and 24 times the rms noise of the map, 0.5~m\jpb. 
The plus sign indicates the position of the Two Micron All Sky Survey (2MASS) point source. 
}   
\label{fcont}
\end{center}
\end{figure}

\begin{figure}
\begin{center}
\begin{tabular}[b]{c}
     \epsfig{file=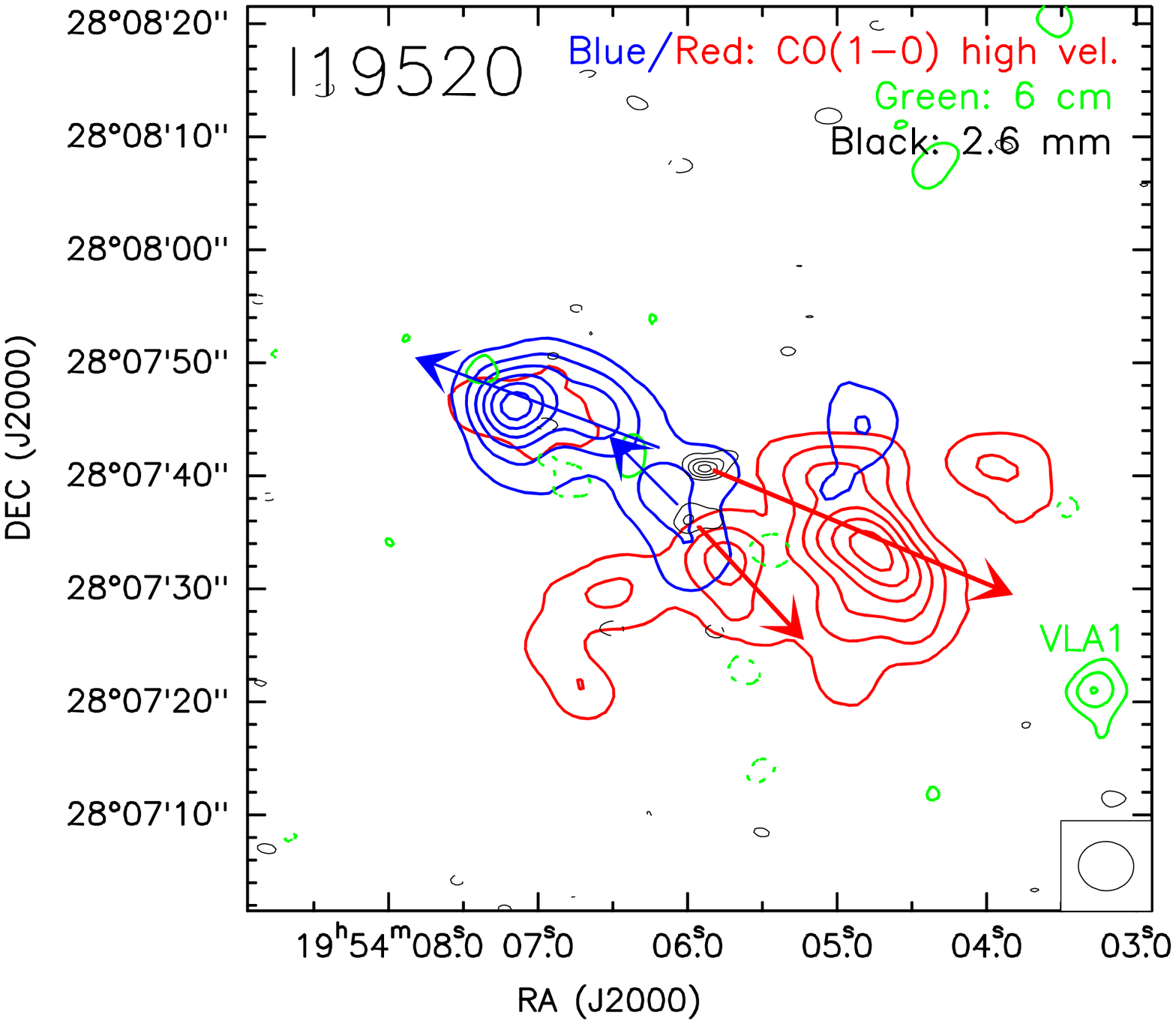,  width=8.2cm, angle=0} \\
\end{tabular}
\caption{
Blue/red contours: OVRO \co\,(1--0) high velocity emission. Blue contours correspond to integration in the velocity interval $-50$ to $-26$~\kms, and red contours correspond to integration in the velocity $-5.8$ to $9.8$~\kms. Blue(red) contours start at 0.15 times the peak intensity, of
13.02(8.18)~\jpb\,\kms, and increase in steps of 0.15 times the peak intensity.
The synthesized beam of the \co\,(1--0) emission, of $4.91\times4.34$~arcsec$^2$, P.A.=85.5\degr, is shown in the bottom-right corner.
Green contours: VLA 6~cm emission. Contours are  $-3$, 3, 6, and 9 times 0.11~m\jpb. The VLA synthesized beam is $3.84\times3.54$~arcsec$^2$ with PA=88.8\degr.
Black contours correspond to the 2.6~mm continuum emission as in Fig.~\ref{fcont}.
}   
\label{fcohv}
\end{center}
\end{figure}

\subsection{VLA archive data \label{sovla}}

We searched the VLA\footnote{The Very Large Array (VLA) is operated by the
National Radio Astronomy Observatory (NRAO), a facility of the National Science
Foundation operated under cooperative agreement by Associated Universities,
Inc.} archive for centimetre continuum observations toward I19520. I19520 was
observed at 6~cm (4.8~GHz) on 27 November 2003 with the array in the B
configuration (project AH832, Urquhart \et\ 2009\footnote{The results of project AH832 are published in Urquhart \et\ (2009), who report no detection in the field of \ii, with a rms noise of 0.16~m\jpb. However, in the cleaning process Urquhart \et\ (2009) used a robust parameter of 0 and did not apply any tapering to the data. Thus, with the aim of recovering possible faint and extended structure we re-did the calibration and imaging using a robust parameter of 5 (to improve the sensitivity) and tapering the data.}),
and at 2~cm (14.9~GHz) on 17 September 1986 with the array in the CnB configuration (project AP121). 
For project AH832, the phase center of the observations was RA\,(2000)=19$^{\rm h}$54$^{\rm m}$05$\fs$90 and
Dec\,(J2000)=28$\degr$07$\arcmin$41$\farcs0$. The absolute flux scale was set by
observing the quasar 1331+305 (3C286), for which we adopted a flux of 7.5~Jy at
6~cm. The quasar 2048+431, with a bootstrapped flux of 0.42~Jy, was observed
regularly to calibrate the gains and phases. The total on-source time was $\sim$2.5~minutes. 
For project AP121, the phase center was RA\,(2000)=19$^{\rm h}$54$^{\rm m}$05$\fs$40 and
Dec\,(J2000)=28$\degr$07$\arcmin$37$\farcs$6. The absolute flux scale was set by
observing 3C286, with and adopted flux of 3.5~Jy at 2~cm; while the quasar
1923+210, with a bootstrapped flux of 1.8~Jy, was used to calibrate the gains
and phases. The source was observed for $\sim$3.5~minutes. 
In both cases, the data reduction followed the VLA standard guidelines for calibration of
high-frequency data (or high-angular resolution data), using the NRAO package
AIPS. Final images were produced with the robust parameter of Briggs (1995) set
to 5, corresponding to natural weighting, and applying a taper at 50~k$\lambda$
with the aim of recovering faint extended emission. The final 1$\sigma$ rms noise levels are
0.1~mJy~beam$^{-1}$ at 6~cm, and 0.6~mJy~beam$^{-1}$ at 2~cm. The synthesized
beams of the resulting images are $3\farcs8\times3\farcs5$ with PA=89\degr\ at
6~cm, and $3\farcs4\times3\farcs0$ with PA=$-$82\degr\ at 2~cm. 

\begin{figure*}
\begin{center}
\begin{tabular}[b]{c}
     \epsfig{file=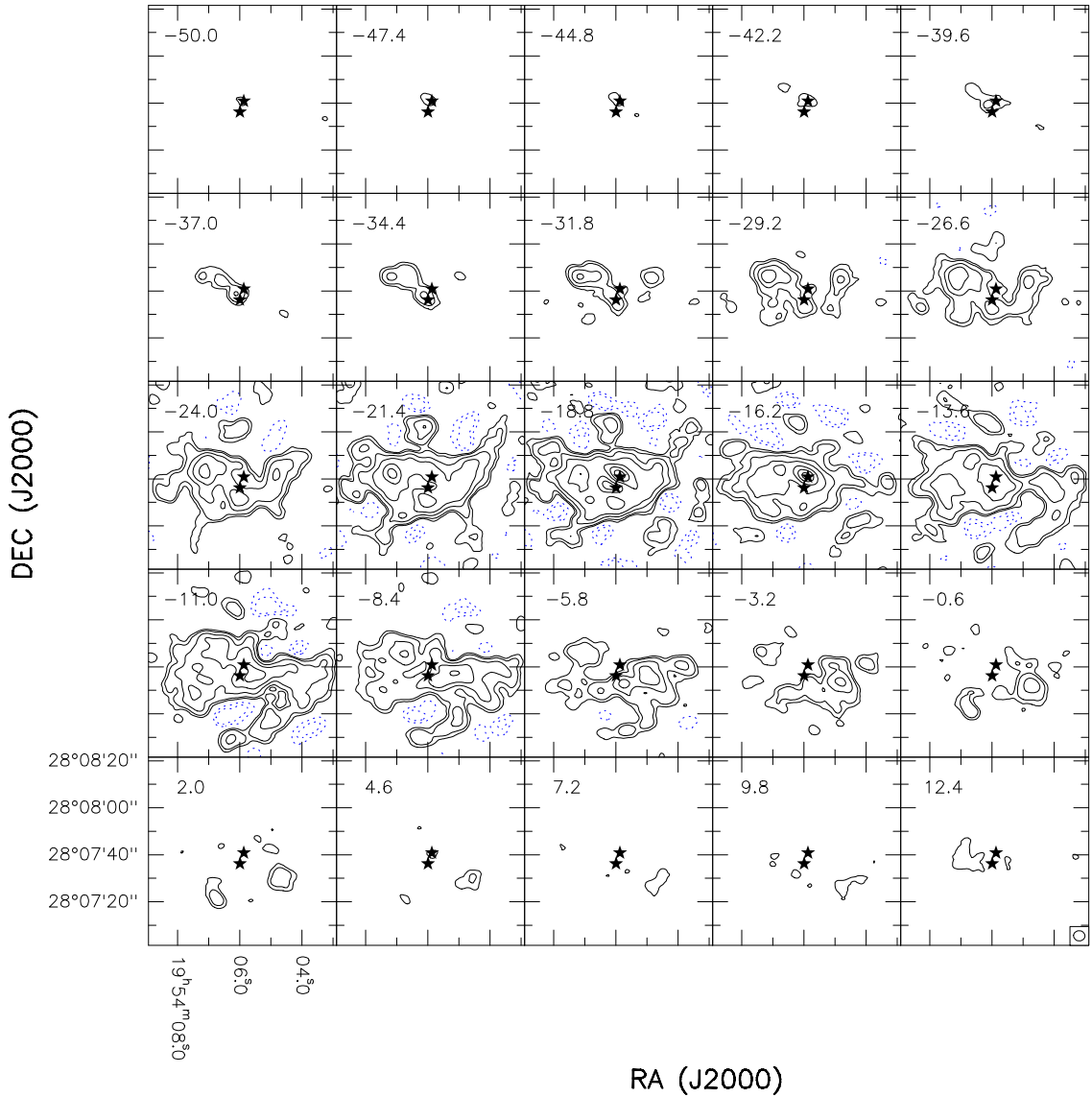 , width=15cm} \\
\end{tabular}
\caption{OVRO \co\,(1--0) channel maps for velocities ranging from $-50$ to 12~\kms. Contours are $-10$, $-5$, 5, 10, 25, 50, 100, 150, 200, 250, and 300 times the rms noise of the map, 0.017~\jpb. The beam is shown in the bottom-right corner of the bottom-right panel, and the 2 stars mark the position of MM1 and MM2.
}   
\label{fcoch}
\end{center}
\end{figure*}

\section{Results}\label{res}

\subsection{OVRO 2.6~mm and VLA 6~cm continuum}\label{srcont}

The 2.6 mm continuum emission toward \ii, after combining data from the 3 OVRO configurations, reveals one source in the field of view, which is slightly resolved, and whose peak position is coincident with the position of the infrared source (see Fig.~\ref{fcont} and Table~\ref{tcont}). Imaging using only the H configuration data (providing the highest angular resolution) shows that the continuum emission splits up into at least two unresolved sources: one strong source detected at 14$\sigma$ (MM1) and one fainter source (MM2) about 4~arcsec to the south detected at $6\sigma$, which has an elongation to the west (MM2W, Fig.~\ref{fcont}).  The parameters of the millimetre sources are listed in Table~\ref{tcont}. For MM1, we measured a deconvolved size of $22000\times6000$~AU, at PA=$-72$\degr.
As for MM2, a double-gaussian fit for MM2 and MM2W was required to obtain a residual map with no excess of emission to the south of MM1, and both sources are unresolved.


The total mass $M$ of gas and dust from thermal continuum emission, assuming that the emission is optically thin, is:

\begin{equation}
M=\frac{S_\nu D^2}{B_\nu(\Td)\kappa_\nu},
\end{equation}
where $S_\nu$ is the flux density at the frequency $\nu$, $D$ is the distance to the Sun, $B_\nu(\Td)$ is the Planck function at the dust temperature $\Td$, and $\kappa_\nu$ is the absorption coefficient per unit of total (gas+dust) mass density. Writing Eq.~(1) in practical units:

\begin{equation}
\left[\frac{M}{M_{\odot}}\right]=3.25\times
\frac{e^{0.048\,\nu/\Td}-1}{\nu^3 \kappa_\nu}\times
\left[\frac{S_\nu}{\mathrm{Jy}}\right]
\left[\frac{D}{\mathrm{pc}}\right]^{2},
\end{equation}
where $\Td$ is in K, $\nu$ is in GHz, and $\kappa_\nu$ is in cm$^2$g$^{-1}$. 
For the absorption coefficient at 115.27~GHz we used the extrapolated value from the tables of Ossenkopf \& Henning (1994), for the case of thin ice mantles and density of 10$^6$~\cmt, of 0.003~cm$^2$\,g$^{-1}$.
Assuming $\Td$ of 50~K,
the total mass estimated from the 2.6~mm continuum emission is around 100~\mo\ for MM1, and 50, and 20~\mo\ for MM2 and MM2W, respectively. The global properties of the millimetre source (from the C+E+H configuration) are listed in the last row of Table~\ref{tcont} and, by adopting a mean molecular weight of 2.8~m$_\mathrm{H}$, and a size (radius) of 13500~AU, we obatined an average density (of H$_2$ molecules) of $\sim1.5\times10^6$~\cmt, H$_2$ column density of $\sim4\times10^{23}$~cm$^{-2}$, and $A_\mathrm{v}\sim440$~mag\footnote{Note that among the 440~mag of visual extinction (derived from millimetre continuum data), only the half ($\sim200$~mag) are affecting the stellar light emitted towards the observer.} (following Frerking, Langer, \& Wilson 1982).

\begin{figure}
\begin{center}
\begin{tabular}[b]{c}
     \epsfig{file=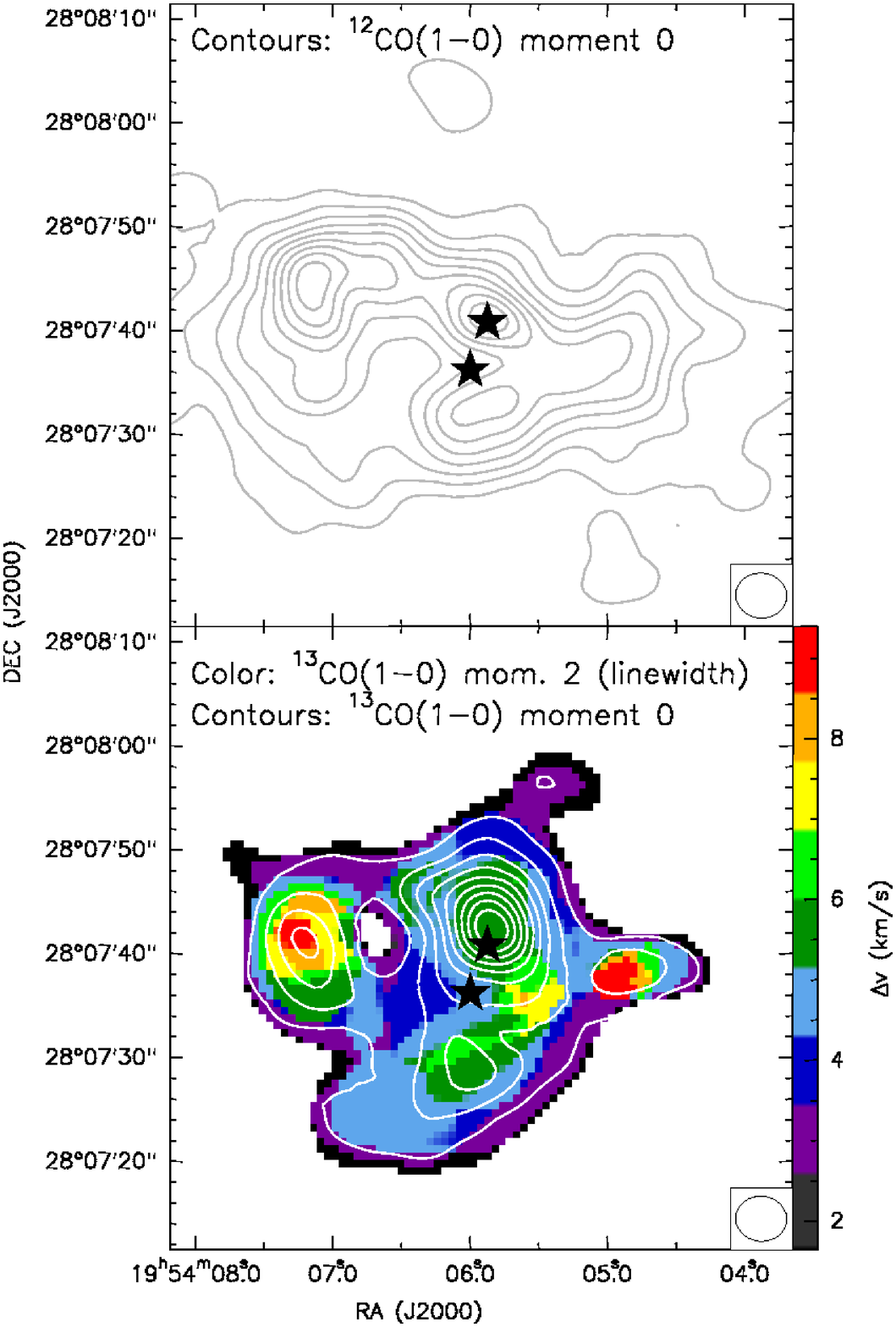, width=8cm} \\
\end{tabular}
\caption{
{\bf Top:} Contours: zero-order moment (integrated intensity over all the velocity range) of \co(1--0) towards \ii. Contour levels range from 10 to 100\% of the peak intensity, 37.07~\jpb\,\kms, increasing in steps of 10\%. 
{\bf Bottom:} Contours: zero-order moment of \tco(1--0) towards \ii. Contour levels range from 10 to 100\% of the peak intensity, 11.63~\jpb\,\kms, increasing in steps of 10\%.
Colorscale: second-order moment (velocity dispersion, which has already been converted to linewidth) of \tco(1--0).
In both panels, star symbols correspond to the position of MM1 and MM2 (Table~\ref{tcont}), and the synthesized beam is shown in the bottom-right corner.
}   
\label{fco-mom0}
\end{center}
\end{figure}

\begin{figure}
\begin{center}
\begin{tabular}[b]{c}
     \epsfig{file=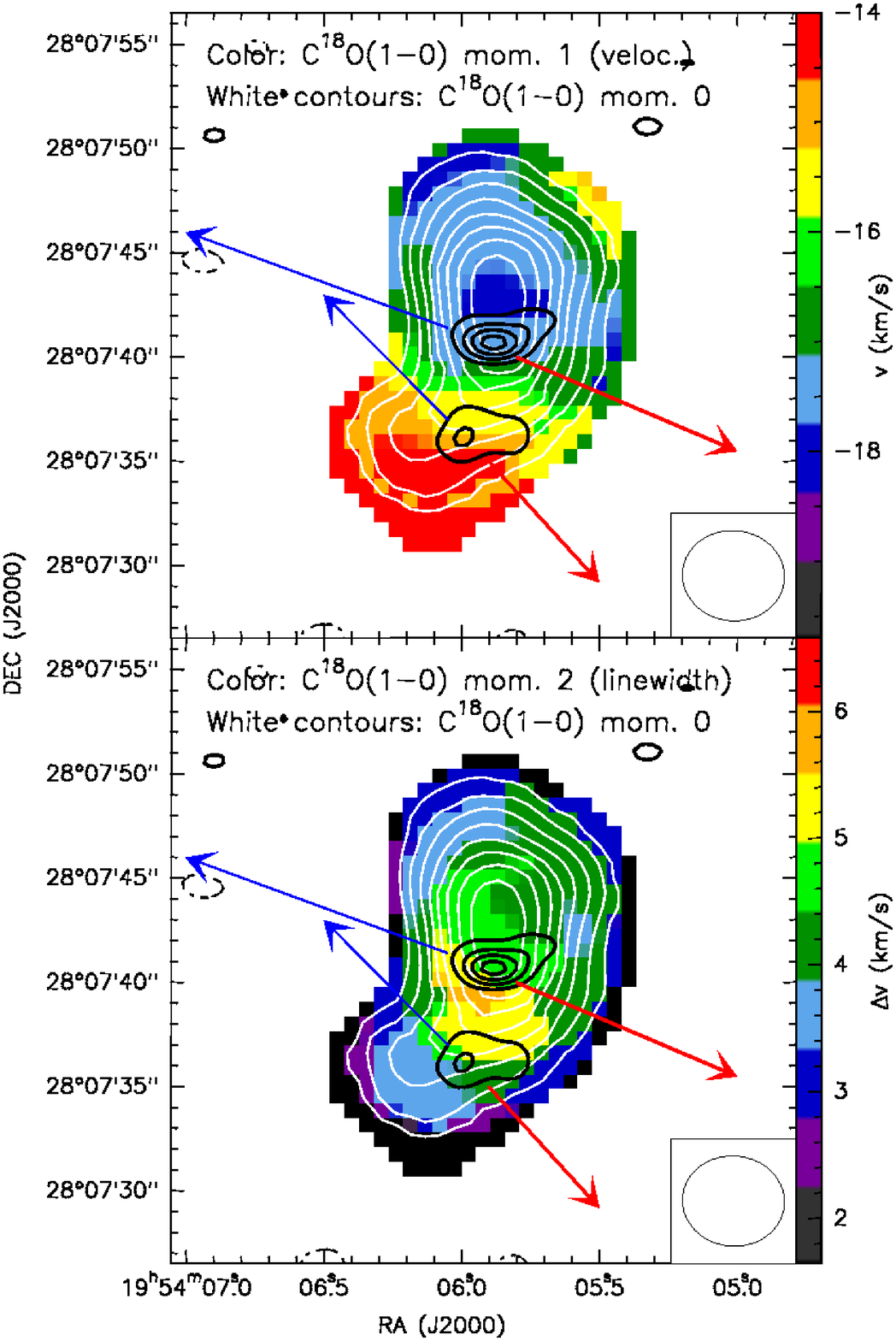, width=7.5cm} \\
\end{tabular}
\caption{
{\bf Top:} White contours: zero-order moment (integrated intensity) of \ceo(1--0) towards \ii. Contour levels range from 20 to 100\% of the peak intensity, 1.54~\jpb\,\kms, increasing in steps of 10\% (emission was integrated over all the velocity range). Colorscale: first-order moment (velocity) of \ceo(1--0).
{\bf Bottom:} Idem as top with the colorscale showing the second-order moment (velocity dispersion, which has already been converted to linewidth).
In both panels, black contours correspond to the H configuration 2.6~mm continuum emission as in Fig.~\ref{fcont}, and the beam is shown in the bottom-right corner.
}   
\label{fc18o}
\end{center}
\end{figure}

Concerning the VLA centimetre emission, we only detected one unresolved source at 9$\sigma$ about 40~arcsec to the southwest of MM1, as shown in Fig.~\ref{fcohv}. The flux density of this source (labeled `VLA1') is $\sim1$~mJy. At 2~cm we detected no source in the field shown in Fig.~\ref{fcohv}, preventing us from drawing any conclusion on the nature of VLA1.
We also downloaded the most recent image at 6~cm of \ii, taken in 2008 within the framework of the CORNISH\footnote{http://www.ast.leeds.ac.uk/cornish/public/index.php} survey (Hoare \et\ 2012). In the \ii\ region, the rms of the 6~cm image is 0.2~m\jpb, and we confirmed again the absence of any strong centimetre source associated with \ii.

\subsection{OVRO \co, \tco\ and \ceo\  emission}\label{CO}

Fig.~\ref{fcoch} shows the channel map of the \co(1--0) emission, spanning 60~\kms\ in total. The emission at high velocities is compact, mainly found to the northeast or to the southwest of MM1, while the emission around  the systemic velocity ($-16.5$~\kms) is much more complex and extends about $>1$~arcmin, being mainly elongated in the east-west direction. In particular, channels at $-11.0$ and $-8.4$~\kms\ show an X-shape structure to the west of MM1 which could be tracing the cavity walls of an outflow.

We computed the zero-order moment (integrated intensity) of the \co\,(1--0) emission for the entire velocity range (Fig.~\ref{fco-mom0}-top), which reveals three main \co\ clumps surrounded by extended emission: one associated with MM1, one about $\sim7$~arcsec to the south, and the other $\ga10$~arcsec to east (there is also a possible fourth clump to the west of MM1). These three clumps are well separated in the zero-order moment of the \tco(1--0) emission, presented in Fig.~\ref{fco-mom0}-bottom. The \tco\ moment-zero map (integrated over all the channels where \tco\ is detected, from $-25$ to $-11$~\kms) indicates that the strongest clump is the one associated with MM1, which was not obvious in the \co\ map probably due to opacity effects. The other fainter \tco\ clumps show however the broadest linewidths, of up to 10~\kms.

We additionally computed the zero-order moment for \co\ in the high velocity range, integrating from $-50$ to $-26$~\kms\ for the blueshifted emission, and from $-5.8$ to 9.8~\kms\ for the redshifted emission, and the result is presented in Fig.~\ref{fcohv}. The figure shows that the emission is bipolar and elongated roughly in the east-west direction, consistent with a bipolar flow driven by MM1. Note that to the south of MM1 there is one blueshifted and one redshifted lobe to each side of MM2, suggesting that MM2 is driving also an outflow elongated in the southwest-northeast direction. 
Since the clumps detected in \tco\  to the south, east and west of MM1 span a broad velocity range, they could be associated with outflow emission as well. 
It is worth noting that the polarized reflection nebula detected by Gledhill (2005) is located about 2~arcsec to the east of \ii, where the blueshifted high-velocity \co\ lobe is located, indicating that possibly the CO bipolar outflow is creating a cavity which is seen in polarized light. This is found in other YSOs where outflows have excavated bipolar cavities (\eg\ Beckford \et\ 2008).


\begin{figure}
\begin{center}
\begin{tabular}[b]{c}
     \epsfig{file=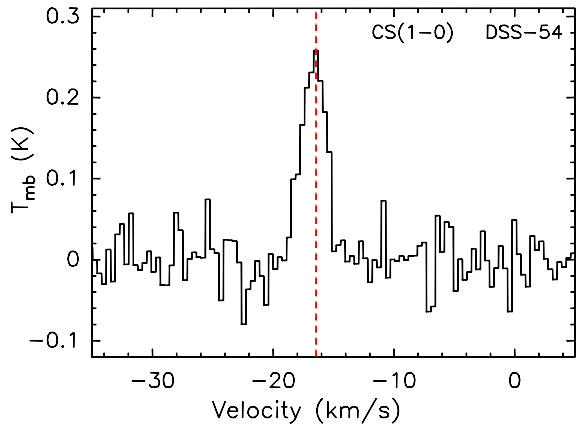,  width=7.5cm, angle=0} \\
\end{tabular}
\caption{CS\,(1--0) line observed with {\bf the DSS-54 antenna of the Madrid DSCC} towards \ii. The spectrum has been smoothed to a spectral resolution of 0.38~\kms.
The dashed line indicates the velocity where the line peaks, $-16.45$~\kms.
}   
\label{fcs}
\end{center}
\end{figure}

Finally, the \ceo\,(1--0) emission, detected in the same velocities approximately as \tco, is concentrated in one clump centred on MM1, with no clumpy emission in its surroundings. The zero-, first- (velocity field), and second-order (linewidth) moments for \ceo\ are shown in Fig.~\ref{fc18o}. 
The integrated \ceo\ emission is elongated roughly in the north-south direction, with the southern part slightly tilted to the east, suggesting that we are observing the superposition of two cores (one north-south, and another one smaller and elongated in the southeast-northwest direction). It is interesting to note that MM1 and MM2 lie at the center of these two \ceo\ cores, and that the elongation of each of the two \ceo\ cores is perpendicular to the corresponding outflows. The velocity field of \ceo\ reveals that MM1 and MM2 are shifted by {\bf $\sim2$~\kms}.

\begin{figure}
\begin{center}
\begin{tabular}[b]{c}
     \epsfig{file=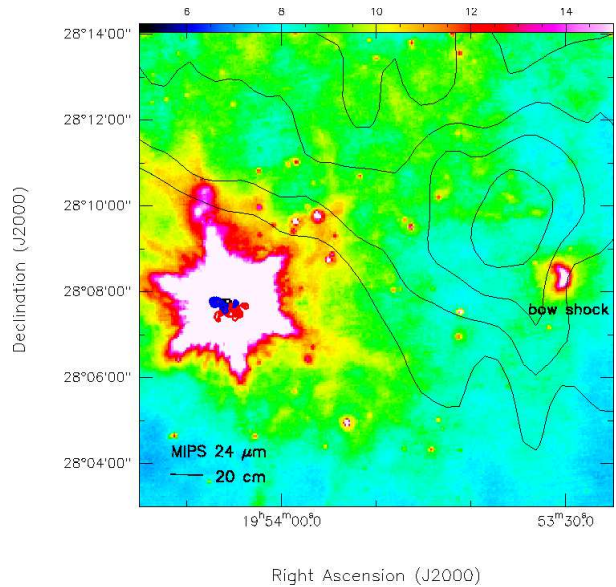,  width=8cm, angle=0} \\
\end{tabular}
\caption{Same as Fig.~\ref{flargefov} (colorscale: 24~\mum\ emission; black contours: 20~cm emission) with the OVRO \co\ high-velocity emission overlaid (red and blue contours). Note the bow shock structure 7~arcmin to the west of \ii, whose tail is pointing towards \ii.
}   
\label{fmips1}
\end{center}
\end{figure}

\begin{figure}
\begin{center}
\begin{tabular}[b]{c}
     \epsfig{file=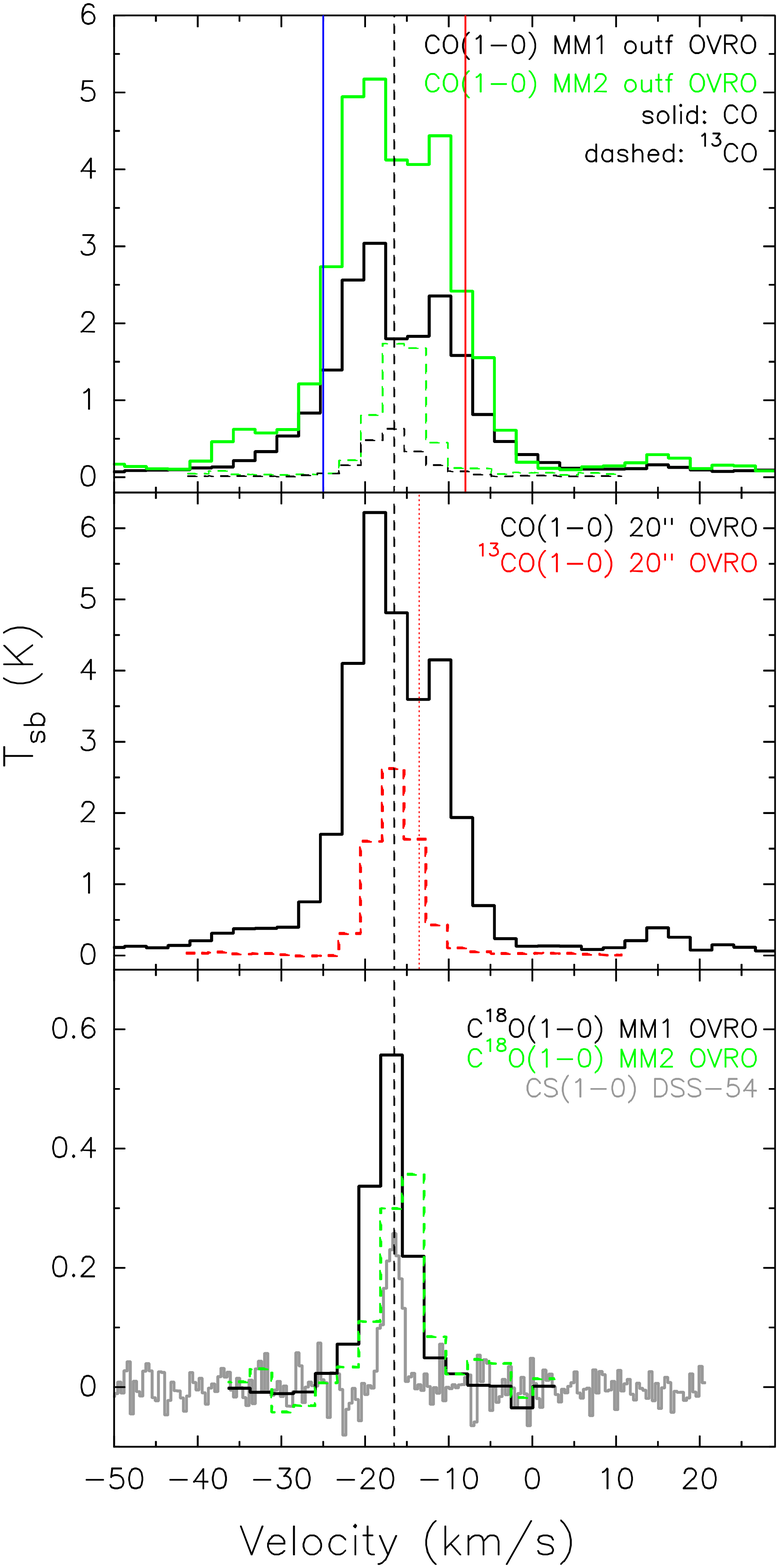, width=6cm, angle=0} \\
\end{tabular}
\caption{
{\bf Top:} Black (solid/dashed): \co\,(1--0)/\tco(1--0) spectrum averaged over a region of $75\times25$~arcsec$^2$ centred on MM1, used to derive the outflow parameters. Green: idem for MM2, for a region of $20\times10$~arcsec$^2$. Blue and red vertical lines indicate where the outflow wing velocity range starts. 
{\bf Middle:} Black (solid): \co\,(1--0) averaged over a region of 20~arcsec (of diameter) centred on MM1. Red (long-dashed): idem for \tco(1--0). The vertical red dotted line indicates the velocity where the dip in \co\ is seen.
{\bf Bottom:} Black (solid): \ceo(1--0) averaged over a region of $15\times10$ arcsec$^2$ centred on MM1. Green (dashed): \ceo(1--0) averaged over a region of $10\times7$ arcsec$^2$ centred on MM2. Grey (solid): CS\,(1--0) spectrum as in Fig.~\ref{fcs} shown for comparison.
In all panels,  the vertical black dashed line indicates the systemic velocity.
}   
\label{fcospec}
\end{center}
\end{figure}

\subsection{CS\,(1--0) with DSS-54 of Madrid DSCC \label{srcs}}

The CS\,(1--0) line was detected using the DSS-54 antenna of Madrid DSCC (see Fig.~\ref{fcs}), and the results of a Gaussian fit to the line are listed in Table~\ref{tisotop}. The linewidth of the line is $\sim2.3$~\kms, which is completely dominated by non-thermal motions, even if we assume that the dense core is at 50~K. This broad linewidth could be produced either by turbulence injected by the outflow or by systemic motions such as rotation and/or infall.
In fact, the line profile is asymmetric, with the blueshifted side being slightly stronger than the redshifted side. This is the expected profile for an optically thick line tracing an infalling envelope (\eg\ Torrelles \et\ 1995; Tsamis \et\ 2008).

\subsection{Spitzer at 24~\mum\ \label{srmips1}}

We downloaded the Spitzer (Infrared Array Camera (IRAC) + MIPS) images of the \ii\ field of view from the MIPSGAL (Carey \et\ 2009) database\footnote{http://irsa.ipac.caltech.edu/data/SPITZER/MIPSGAL/}. In Fig.~\ref{fmips1} we show a superposition of the \co\  high-velocity OVRO emission on the 24~\mum\ colorscale of Spitzer/MIPS. The MIPS image reveals a structure with a bow-shock morphology about 7~arcmin to the west of \ii, whose tail is pointing towards \ii, suggesting wind/jet activity in the east-west direction as suggested also by the high-velocity \co\ outflow. Such a wind/jet activity is consistent with the P-Cygni profile seen in the H$\alpha$ line of \ii, from which a maximum expansion velocity between 400 and 800~\kms\ is inferred (S\'anchez Contreras \et\ 2008). Using an average velocity of 600~\kms, we estimate that the ejection event producing the bow shock took place about $\sim30000$\,tan($i$)~yr ago (with $i$ the inclination with respect to the plane of the sky).


\section{Analysis}

\subsection{Physical parameters from \co, \tco, \ceo\ for MM1 and MM2}

In Fig.~\ref{fcospec} we show the \co\,(1--0), \tco\,(1--0), and \ceo\,(1--0) spectra averaged in different regions. 
The results from a Gaussian fit for all but the \co\ transition are listed in Table~\ref{tisotop} (corresponding to the spectra in the middle and bottom panels of Fig.~\ref{fcospec}), together with the derived opacities, $\tau$, and column densities, $N$.
Opacities for \tco\ and \ceo\ were estimated from equation:

\begin{equation}
\tau=-\mathrm{ln}\left[1-\frac{T_\mathrm{L}}{J_\nu(\Tex)-J_\nu(\Tbg)}\right],
\end{equation}
where $T_\mathrm{L}$ is the intensity at the line center,  $\Tex$ is the excitation temperature, $\Tbg$ is the background tamperature (adopted 2.7~K), and $J_\nu(T)=h\nu/k/(e^{h\nu/kT}-1)$ (with $k$ being the Boltzmann constant, $T$ the temperature and $\nu$ the frequency). $\Tex$, of around 9~K, was estimated from \co\ assuming optically thick emission, and was found to be slightly smaller than the temperature obtained by Arquilla \& Kwok (1987, 13~K). This difference is reasonable if we take into account the flux calibration uncertainties and the fact that Arquilla \& Kwok (1987) observed with a single-dish antenna (with a HPBW of 45~arcsec) while we used the OVRO interferometer, being thus sensitive to different spatial scales (see below).

We derived the \co\ column density (from the 20~arcsec averaged spectrum) following Palau \et\ (2007), and using the opacity derived from \tco\,(1--0). The \tco\,(1--0) column density was estimated with the equation:

{\footnotesize
\begin{equation}
\left[\frac{N_\mathrm{isotp}}{\mathrm{cm^{-2}}}\right]=2.48 \times 10^{14}
\left[\frac{\Tex}{K}\right]\,
\left[\frac{\Delta v}{\mathrm{km~s^{-1}}}\right]\,
\frac{\tau}{1-e^{h\nu/(k\Tex)}},
\end{equation}
}
\noindent which can be used for \ceo(1--0) as well. The obtained column densities were around $2\times10^{16}$~cm$^{-2}$ for \tco(1--0), and (2--3)$\times10^{15}$~cm$^{-2}$ for \ceo(1--0).
As for the mass, we used the size listed in Table~\ref{tisotop}, adopted a mean molecular weight of 2.8, and abundances of $X$(\co)=$10^{-4}$, $X$(\tco)=$10^{-6}$, and $X$(\ceo)=$1.7\times10^{-7}$ (Solomon, Sanders, \& Scoville \et\ 1979; Frerking, Langer, \& Wilson 1982; Scoville \et\ 1986). The final masses are around 80~\mo\ including both MM1 and MM2 (from \tco), and about 40~\mo\ and 20~\mo\ for MM1, and MM2, respectively (from \ceo, for which we could separate the emission from MM1 and MM2). 

The systemic velocity is found at $\sim -16.5$~\kms\ (measured from \tco\ and CS), with MM2 being about 2~\kms\ redshifted with respect to MM1. The velocity is similar to that derived from \water\ and OH maser (Engels \et\ 1984; Lewis \et\ 1985).
%
The \co\ spectrum (Fig.~\ref{fcospec}-middle) shows a double-peaked profile which could be due in part to the filtering of emission at systemic velocities and/or self-absorption of the cold foreground cloud. This is expected because \co\,(1--0) is optically thick and traces gas of relatively low density ($\la1000$~\cmt). However, for opacity/filtering effects one would expect to see the maximum absorption at the systemic velocity, while this is not observed. In fact, the \co\ profile could be well explained if the material is infalling towards the central object, as the blueshifted \co\ peak (and wing) is stronger than the redshifted one, and the dip is redshifted by about 3~\kms\ with respect to the systemic velocity. 
Using the infall velocity derived from the \co\ absorption dip, an approximate range of radii associated with this infall velocity (3000--36000~AU, from the measured size of the millimetre source and the \ceo\ clump, Tables~\ref{tcont} and \ref{tisotop}), and the H$_2$ column density derived from \tco\ (Table~\ref{tisotop}), we estimated, following Beltr\'an \et\ (2006), a mass infall rate of (0.8--2.6)$\times10^{-4}$~\mo\,yr$^{-1}$.

\begin{table*}
\caption{Physical parameters from averaged OVRO \tco\,(1--0), and \ceo\,(1--0) spectra, and DSS-54 single-pointing CS\,(1--0)}
\centering
\footnotesize
\begin{tabular}{lccccccccc}
\hline\hline\noalign{\smallskip}
&$T_\mathrm{peak}$
&$v$
&$\Delta v$
&Area
&
&$N_\mathrm{isotp}$
&$N_\mathrm{H2}$\supb
&Size\supc
&Mass
\\
Transition
&(K)
&(\kms)
&(\kms)
&(K\,\kms)
&$\tau$\supa
&(cm$^{-2}$)
&(cm$^{-2}$)
&(arcsec)
&(\mo)
\\
\noalign{\smallskip}
\hline\noalign{\smallskip}
\tco(1--0) MM1+MM2	&$2.61\pm0.01$	&$-16.54\pm0.02$	&$6.1\pm0.1$	&$17.1\pm0.1$	&0.56	&$1.8\times10^{16}$ &$1.8\times10^{22}$ 	&11	&75\\
\ceo(1--0) MM1		&$0.56\pm0.02$	&$-17.20\pm0.08$	&$5.5\pm0.2$	&$3.3\pm0.1$	&0.10 	&$2.9\times10^{15}$ &$1.7\times10^{22}$	&8	&36\\
\ceo(1--0) MM2		&$0.38\pm0.03$	&$-15.37\pm0.19$	&$5.5\pm0.5$	&$2.2\pm0.2$	&0.06 	&$1.9\times10^{15}$ &$1.1\times10^{22}$	&7	&18\\
CS\,(1--0) MM1+MM2	&$0.26\pm0.03$	&$-16.65\pm0.07$	&$2.3\pm0.2$	&$0.64\pm0.04$	&$-$ 	&$-$				&$-$&$-$ &$-$\\
\hline
\end{tabular}
\begin{list}{}{}
\item[$^\mathrm{a}$] Opacity estimated from the intensity at the line center, and using an excitation temperature of 9.3~K (derived from \co\ for the optically thick case). 
\item[$^\mathrm{b}$] The adopted abundances are X(\tco)=$10^{-6}$ (Solomon, Sanders, \& Scoville 1979), and X(\ceo)=$1.7\times10^{-7}$ (Frerking, Langer, \& Wilson 1982).
\item[$^\mathrm{c}$] FWHM. 
\end{list}
\label{tisotop}
\end{table*}

\begin{table*}
\caption{Physical parameters of the outflows driven by MM1 and MM2}
\centering
\footnotesize
\begin{tabular}{lccccccccc}
\hline\hline\noalign{\smallskip}
&$t_\mathrm{dyn}$
&size
&$N_{12}$~$^\mathrm{a}$
&$M_\mathrm{out}$~$^\mathrm{a}$
&$\dot{M}$~$^\mathrm{a}$
&$P$~$^\mathrm{a}$
&$\dot{P}$~$^\mathrm{a}$
&$E_\mathrm{kin}$~$^\mathrm{a}$
&$L_\mathrm{mech}$~$^\mathrm{a}$
\\
Lobe
&(yr)
&(arcsec)
&(cm$^{-2}$)
&(\mo)
&(\mo~yr$^{-1}$)
&(\mo~\kms)
&(\mo~\kms~yr$^{-1}$)
&(erg)
&(\lo)
\\
\noalign{\smallskip}
\hline\noalign{\smallskip}
MM1-Red	&39000	&$23\times9$	&$6.1\times10^{15}$	&0.54	&$1.4\times10^{-5}$&13	&$3.3\times10^{-4}$	&$3.1\times10^{45}$	&0.45\\
MM1-Blue	&24000	&$20\times8$	&$5.5\times10^{15}$	&0.38	&$1.5\times10^{-5}$&13	&$5.5\times10^{-4}$	&$4.7\times10^{45}$	&1.1\\
MM1-All	&32000	&-      		&$1.2\times10^{16}$	&0.92	&$2.9\times10^{-5}$&26	&$8.3\times10^{-4}$	&$7.8\times10^{45}$	&1.6\\
\hline\noalign{\smallskip}
MM2-Red	&20000	&$7\times5$	&$1.9\times10^{16}$	&0.28	&$1.4\times10^{-5}$&4	&$1.8\times10^{-4}$	&$4.5\times10^{44}$	&0.11\\
MM2-Blue	&9000	&$4\times4$	&$3.0\times10^{16}$	&0.21	&$2.3\times10^{-5}$&3	&$2.9\times10^{-4}$	&$3.3\times10^{44}$	&0.14\\
MM2-All	&14000	&-      		&$4.9\times10^{16}$	&0.49	&$3.4\times10^{-5}$&6	&$4.2\times10^{-4}$	&$7.8\times10^{44}$	&0.26\\
\hline
\end{tabular}
\begin{list}{}{}
\item[$^\mathrm{a}$] Parameters are calculated following Palau \et\ (2007), for an inclination with respect to the plane of the sky equal to 45\degr, and are corrected for opacity using the \tco\ spectrum (see main text). For MM1 we estimated an intensity in the line wings of \tco\ of about 0.05~K (implying an opacity of \co\ of 0.02 and a correction of $\sim2$). From the line peak of the \co\ spectrum, $\sim3$~K, we estimated and excitation temperature of $\sim6$~K, assuming optically thick emission.
For MM2 we estimated an intensity in the line wings of \tco\ of about 0.1~K (implying an opacity of \co\ of 0.03 and a correction of $\sim3$). From the line peak of the \co\ spectrum, $\sim5$~K, we estimated an excitation temperature of $\sim8$~K, assuming optically thick emission. See Section~\ref{saoutfpar} for further details.
\end{list}
\label{toutfpar}
\end{table*}

\subsubsection{Parameters of the \co\ outflows \label{saoutfpar}}

Table~\ref{toutfpar} lists the parameters estimated for the MM1 and MM2  \co(1--0) outflows, computed following Palau \et\ (2007; see also notes in the table; we used the \co\ spectra shown in Fig.~\ref{fcospec}-top) and correcting for opacity effects.  
For the outflow inclination (with respect to the plane of the sky), $i$, we adopted an intermediate angle of 45\degr, consistent with the bipolar outflow morphology.

The dynamical times of the MM1 and MM2 outflows are in the range 14000--32000~yr. In particular, the dynamical time of the outflow driven by MM1, of around 32000~yr, is very similar to the estimated timescale for the ejection event which produced the bow shock seen at 24~\mum\ about 7~arcmin to the west of \ii\ (see Fig.~\ref{fmips1} and Section~\ref{srmips1}), reinforcing that \ii\ probably had a recent and powerful ejection event. In fact, if the bow-shock is indeed driven by \ii, this would be one of the longest jets known to date, extending for $18/$cos($i$)~pc, similar to or even larger than the recently published updated length of the HH80-81 jet (Masqu\'e \et\ 2012).
As for the other outflow parameters, they are slightly smaller than the single-dish outflow parameters derived by Arquilla \& Kwok (1987, see their  Table~2). This was expected because the interferometer filters out the most extended emission. For the case of our OVRO observations, using the minimum $uv$-distance (see Section~2) and following Palau \et\ (2010), we estimated that OVRO was filtering structures larger than $\sim23''$.
From the \co(1--0) single-dish observations, Arquilla \& Kwok (1987) estimate a total flux integrated over all the velocity range (single-point observations) of 1890 Jy~km/s, and for the OVRO data we estimate (integrating both spectrally and spatially) 870~Jy~\kms. Thus OVRO  is recovering 46\% of the single-dish flux, or missing a flux of 54\%.

In order to estimate the collimation factor of the outflow driven by MM1 (\ie\ excluding the outflow lobes of MM2), we followed the method used in Bally \& Lada (1983) and  Wu \et\ (2004), where the collimation factor is  the ratio of the major to minor radii of the outflow considered as an ellipsoid. By fitting a Gaussian to the redshifted lobe and another Gaussian to the blueshifted lobe, we obtained the minimum radius, of about 7~arcsec (deconvolved minor size of the gaussian fit), and the maximum radius was estimated from the total length of $\sim40$~arcsec. This yields a collimation factor of 5.7, which is among the most high collimation factors measured in outflows driven by very luminous ($\sim10^5$~\lo) YSOs, either using single-dish telescopes (\eg\ Wu \et\ 2004; Qiu \et\ 2011) or interferometer arrays (\eg\ Cesaroni \et\ 1999; Gibb \et\ 2003; Beuther, Schilke, \& Gueth 2004; Kumar, Tafalla \& Bachiller 2004; Brooks \et\ 2007; Zhang \et\ 2007; Qiu \et\ 2007; 2009, 2012; Zapata \et\ 2009a, 2011; Masqu\'e \et\ 2012). 

\begin{figure}
\begin{center}
\begin{tabular}[b]{c}
     \epsfig{file=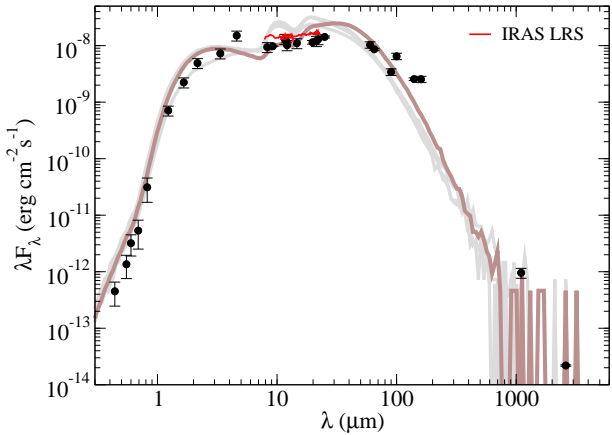,  width=7.2cm, angle=270} \\
\end{tabular}
\caption{Observed and model SED of \ii\ using Robitaille \et\ (2007) tool. The filled circles show the observed fluxes. The thick coloured lines correspond to the best fit model (brown) and the three next best fits (grey). The models are calculated for an aperture of 48000~AU. The red thin line corresponds to the IRAS LRS spectrum.
}   
\label{fsed}
\end{center}
\end{figure}

\subsection{Spectral Energy Distribution \label{sased}}

We built the Spectral Energy Distribution (SED) by compiling data from 2MASS (Skrutskie \et\ 2006), WISE (Wright \et\ 2010), Akari (Murakami \et\ 2007), IRAS (Neugebauer \et\ 1984), Midcourse Space EXperiment (MSX, Egan \et\ 2003), Bolocam (Rosolowsky \et\ 2010, Aguirre \et\ 2011), and OVRO (this work). The global SED, shown in Fig.~\ref{fsed}, has a flat spectrum from 3 to 100~\mum.
In addition, we included optical photometry at 0.6 and 0.8~\mum~extracted
from Hubble Space Telescope images of \ii\footnote{\ii\ was imaged with the ACS/HRC instrument in the F606W and F814W  broad band filters as part of a survey for candidate pre-planetary nebulae (GO 9463: Sahai et al. 2007), and appeared to be starlike in these images.}, and we also downloaded the IRAS Low Resolution Spectrometer (LRS) spectra\footnote{IRAS Science Team (1986); for further deatils on LRS spectra, see http://irsa.ipac.caltech.edu/IRASdocs/exp.sup/; to download the spectrum we used VizieR, Ochsenbein \et\ (2000).} of \ii\ (7.7--22.6~\mum).
The LRS spectrum of \ii\ is classified by Kwok, Volk, \& Bidelman (1997) as `unusual', indicating a flat continuum spectrum with unusual features. The spectrum is quite noisy from 11~\mum\ on, and shows no evidence of any silicate feature, apart from a very faint possible absorption at 9.7~\mum, with a weak peak at shorter wavelengths. A weak silicate absorption with a weak peak at 8.5~\mum\ has also been found in other massive YSOs (\eg\ Chen, Wang, \& He 2000; Campbell \et\ 2006).
New higher sensitivity observations should be carried out to properly study the spectral features of \ii\ in this wavelength range.



We have fitted the \ii\ SED using the online tool provided by Robitaille et al.\,(2007), which computes least-squares fits of
pre-computed models of YSOs having discs and rotationally-flattened infalling envelopes (with biconical outflow cavities), to user-defined SEDs. 
We note that the parameters of the best-fit models of the online tool must be regarded with caution, as recent work confronting the online tool results with spatial information data (\eg\ Linz \et\ 2009; de Wit \et\ 2010, 2011) indicate that the spatial information is required to break the degeneracy inherent to some of the online tool output parameters and to overcome the limited sampling of the parameter space (see Robitaille 2008; Linz \et\ 2009; Offner \et\ 2012). Thus, we use the on-line SED fitting tool of Robitaille \et\ (2007) as a first approach to the main properties of \ii.

We set the input distance range to $D=8.5-9.5$\,kpc. The input interstellar extinction range was set to $A_\mathrm{v}=3-5$, based on our estimate of $A_\mathrm{v}\sim4.1\pm\,0.9$ using the numerical algorithm provided by Hakkila et al. (1997), which computes the 3-dimensional visual interstellar extinction and its error from inputs of Galactic longitude, latitude, and distance, from a synthesis of several published studies.
The best-fit model has D=8.7\,kpc, $A_\mathrm{v}=4.4$, a central star with mass $M_*=29$\,\mo, effective temperature $T_\mathrm{eff}=39990$\,K, luminosity $L=1.2\times10^5$~\lo, 
and age $t_\mathrm{age}=6.1\times10^4$\,yr; a central disc mass $M_\mathrm{disc}=0.3$~\mo, and an envelope mass $M_\mathrm{env}=138$~\mo. In this model, 
the (half) opening angle of the bipolar outflow cavity is $\theta_c=31$\degr, and the outflow cavity axis is inclined at an  
angle of $i=59$\degr\ to the plane of the sky. 
See Table~\ref{tsed} for the ranges of fitted values for the next three best-fit models.
These models provide a good fit of the SED out to the far-infrared wavelengths. But at millimetre wavelengths, the models (based on a Monte Carlo code) do not have enough S/N, hence we are not able to make a comparison of the model and data for these wavelengths. The circumstellar  extinction is $A_\mathrm{v}\mathrm{(csm)}=2.4$ in the best fit model, and lies in the range 3.1--4.1 for the next three best-fit models. This makes a total \Av\ (foreground + circumstellar) of $\sim7$~mag for the best fit model.


\begin{table}
\caption{Results from the SED fit (Robitaille \et\ 2007) }
\begin{center}
{\small
\begin{tabular}{lcc}
\noalign{\smallskip}
\hline\noalign{\smallskip}
&Best fit
&Range of
\\
Parameter
&values
&values\supa\
\\ 
\noalign{\smallskip}
\hline\noalign{\smallskip}
Distance (kpc)						&8.7     			&8.7--9.5\\
$A_\mathrm{v}$\supb\ (mag)			&4.4				&3.0--3.8\\
$M_*$ (\mo)						&29				&26--33\\
$T_\mathrm{eff}$ (K)				&39990			&38650--41500\\
$L_\mathrm{bol}$ (\lo)				&$1.2\times10^5$	&(1.0--1.9)$\times10^5$\\
Age (yr)							&61000			&(1.1--2.2)$\times10^5$\\
$M_\mathrm{disc}$ (\mo)				&0.3				&0.5--2\\
$M_\mathrm{env}$ (\mo)				&138			&52--82\\
Envelope outer radius (AU)			&10$^5$			&10$^5$\\
Outflow opening angle $\theta_c$ (\degr)	&31				&13--27\\
Outflow inclination (\degr)\supc\		&59				&59--72\\
\hline
\end{tabular}
\begin{list}{}{}
\item[\supa] Range of values for the three next best fits.
\item[\supb] Foreground interstellar extinction.
\item[\supc] Outflow inclination with respect to the plane of the sky.
\end{list}
}
\end{center}
\label{tsed}
\end{table}



\section{Discussion}\label{sd}

\subsection{Comparison of the molecular gas and dust envelope properties of IRAS\,19520+2759 with other massive YSOs \label{sdcomp}}

In previous sections we have shown that the OVRO observations reveal a strong dust condensation, referred to as MM1, with a total mass  of around 100~\mo, and which is driving a high-velocity collimated \co\,(1--0) outflow. The parameters of the outflow driven by MM1, if compared to those of outflows driven by high-mass YSOs, fit reasonably well the value expected for a massive YSO of $\sim10^4$~\lo\ (\eg\ Wu \et\ 2004; L\'opez-Sepulcre \et\ 2009), after applying a factor of $\sim2$ to account for the 54\% of flux filtered out by OVRO (the works reporting correlations of outflow parameters use single-dish data and thus are sensitive to the most extended emission, which has been filtered out by OVRO). 

To further compare the \ii\ envelope properties with previous works, we estimated the mass of the cloud associated with \ii\, by using the Bolocam image (Rosolowsky \et\ 2010; Aguirre \et\ 2011, see Fig.~\ref{flargefov}), for which we measured a flux density at 1.1~mm of 0.7~Jy, corresponding to 360--610~\mo, for a dust temperature in the range 20--30~K and using the opacity law of Ossenkopf \& Henning (1994, $\sim0.0137$~cm$^2$\,g$^{-1}$ at 1.1~mm). This cloud mass and the CS\,(1--0) linewidth that we have measured (Table~\ref{tisotop}) are consistent with the correlation found by Beuther \et\ (2002) for a sample of high-mass protostellar objects. 
Similarly, the \tco(1--0) linewidth that we have measured follows the correlation with the bolometric luminosity found by Liu \et\ (2010) for a sample of 98 massive YSOs.

Finally, we plotted the values of  the cloud mass and the measured bolometric luminosity on the mass-luminosity plot of Sridharan \et\ (2002) and Molinari \et\ (2008), and the values follow well the trend for massive YSOs. In particular,  figure 9 of Molinari \et\ (2008) indicates that \ii\ probably is in a transition stage between actively accreting objects and the `envelope clean-up' phase.
This is consistent with the fact that different methods used to estimate \Av\ yield significantly different results, suggesting a non-homogenous density structure in the envelope and possibly the presence of holes and cavities. From the millimetre continuum emission we derived, assuming an homogeneous density distribution,  \Av$\sim200$~mag (Section~\ref{srcont}). This is about one order of magnitude larger than the \Av\ estimated from the fit to the SED (total \Av\ is $\sim7$, Section~\ref{sased}), and from the optical spectrum (ratio of H$\alpha$ to H$\beta$ yields $\sim12$~mag, S\'anchez Contreras, in prep.), which are both along the line of sight. Overall, \ii\ is probably in a phase of main accretion and simultaneous start of envelope disruption as the jet interacts with the envelope creating a cavity. 
Examples of massive YSOs reported in the literature driving outflows and with hints of simultaneous envelope disruption are IRAS\,05506+2414 (Sahai \et\ 2008) and V645\,Cyg (Hamann \& Persson 1989; Clarke \et\ 2006; Miroshnichenko \et\ 2009), with \ii\ being more luminous, and embedded in a one order of magnitude more massive envelope.

\subsection{The central star of IRAS\,19520+2759 \label{sdcentralstar}}

The agreement between the position of the optical/infrared source and the millimetre source suggests that they are tracing the same object. 
The position of the 2MASS source is RA(J2000): 19:54:05.86, Dec (J2000): +28:07:40.6. The position of MM1 (Table~\ref{tcont}) is offset with respect to the 2MASS position by (+0.15, +0.27) arcsec, which falls well within the OVRO ($\sim0.2$~arcsec) and 2MASS ($\sim0.1$~arcsec, Skrutskie \et\ 2006) positional uncertainties.
In addition, the preliminary analysis of the optical spectrum of \ii\ presented in S\'anchez Contreras \et\ (2008) points to a hot object photoionizing the circumstellar gas and driving a dense and compact wind. Thus, the properties of both the optical source and the millimetre source indicate that we are dealing with a young and massive object. 


Further hints on the nature of the optical source may come from the non-detection of centimetre emission: 
if the central star was a (main-sequence) O9 star with a typical ionizing flux $N_\mathrm{i}\sim10^{48}$\,s$^{-1}$ (\eg\ Panagia 1973; Smith, Norris \& Crowther 2002), the expected 6~cm flux density from an optically thin HII region, with an electron temperature of $\sim10000$~K and placed at a distance of 9~kpc, would be $100$~mJy, which should have been well detected with the VLA observations (Section~\ref{sovla})\footnote{If we assume optically thick emission, the 6~cm upper limit can only be reproduced with an O-type star if the free-free emission comes from a region of very small size ($\ll1000$~AU) and large electron density ($\gg10^5$~\cmt), which is very unlikely because such a small size (typical of hyper-compact HII regions, \eg\ Kurtz 2005) would require a very dense envelope strongly infalling and quenching the ionized gas (\eg\ Walmsley 1995; Molinari \et\ 1998), preventing the object to be detected in the optical (\eg\ G10.6$-$0.4: Sollins \& Ho 2005; G28.20$-$0.05: Sollins \et\ 2005; NGC\,7538-IRS1: Sandell \et\ 2009).}.
If the spectral type is B1 (or later), the non-detection of the 6~cm emission is naturally explained, as the flux of ionizing photons in this case is smaller than for the O9 case by one--two orders of magnitude.

\subsection{IRAS\,19520+2759: a B-type star becoming an O-type star?}

As discussed above, the centimetre properties of \ii\ are most consistent with an early-B type classification for the central star; however, 
the spectral type inferred from its high luminosity, $\sim10^5$~\lo, assuming it is a zero-age-main-sequence star, is significantly earlier, O6--O7 (\eg\ Bernasconi \& Maeder 1996; Martins \et\ 2005).
The properties of \ii\ can be reconciled if the object is accreting, which is strongly suggested by several observational facts: 
i) the infrared source is associated with a compact and dusty envelope, MM1, of $\sim100$~\mo; 
ii) MM1 is driving a high-velocity collimated outflow with a dynamical timescale of $\sim30000$~yr;
and iii) the OVRO \ceo(1--0) emission, with a size of $\sim70000$~AU, is elongated perpendicular to the outflow, similar to the molecular toroids found in other massive star-forming regions (\eg\ Beltr\'an \et\ 2005, 2011; Beuther \et\ 2007b; Zapata, Tang, \& Leurini 2010; Furuya \et\ 2011), which usually show a rotation velocity pattern. 
All this suggests that \ii\ has been recently undergoing active accretion.


Hoare \& Franco (2007) have proposed, and numerical simulations have shown (\eg\ Yorke \& Bodenheimer 2008; Hosokawa \& Omukai 2009; Hosokawa, Yorke, \& Omukai 2010), that the main effect of accretion onto a massive star (with mass accretion rates up to 10$^{-3}$~\mo\,yr$^{-1}$), regardless of the geometry of the accretion, is an increase of the stellar radius, giving rise to the so-called `swollen' or `bloated' stars. 
The `swollen-star' scenario naturally explains the low ionizing flux observed in \ii, even though with a bolometric luminosity of $10^5$~\lo. In this scenario, the stellar radius attains its maximum value when the star has accreted $\sim10$~\mo\ (Hosokawa, Yorke, \& Omukai 2010). Assuming that this is the current value of the stellar mass of \ii, and
given the mass measured for the envelope of $\sim100$~\mo\ (Section~3), the total mass which could still be accreted onto the star, for a star formation efficiency of about $\sim10$--20\%,  is 10--20~\mo, implying a final stellar mass of 20--30~\mo, which corresponds to an O-type star. 

The `swollen-star' scenario has also been proposed for other massive YSOs. Davies \et\ (2011) require a `swollen-star' phase to reproduce the observed luminosity distribution of massive YSOs in the sample of the Red MSX Source survey (RMS, \eg\ Urquhart \et\ 2008; see also Mottram \et\ 2011); Simpson \et\ (2012) and Bik \et\ (2012) explain the low effective temperatures or offsets from main-sequence in the Hertzsprung Russell diagram by invoking a `swollen-star' phase for the massive YSOs of  G333.2$-0.4$ or W3\,Main complexes. Finally, individual objects studied in detail within the `swollen-star' scenario are B275 (Ochsendorf \et\ 2011), M8E-IR (Linz \et\ 2009), CRL\,2136 (de Wit \et\ 2011), and Orion\,KL\,IRc2 (Morino \et\ 1998; Testi, Tan, \& Palla 2010). \ii, with a bolometric luminosity around 10$^5$~\lo, would be the most luminous of the four aforementioned `swollen-star' candidates.
Accurate spectral type classification of the central star through, \eg\ optical spectroscopy in the 4000--5000~\AA\ range, is needed to firmly establish the `swollen' star hypothesis proposed for \ii.

\section{Conclusions} \label{conc}

We present the results of OVRO observations at 2.6~mm and \co\,(1--0), \tco\,(1--0), and \ceo\,(1--0) transitions, as well as observations of CS\,(1--0) with the 34\,m antenna DSS-54 of Madrid DSCC, towards IRAS\,19520+2759, a very bright infrared object embedded in a massive cloud. Our main conclusions can be summarized as follows:

\begin{itemize}

\item[-] The 2.6~mm continuum emission is dominated by one strong millimetre source, MM1, clearly associated with the bright infrared source, and which is only barely resolved, with a size of $22000\times6000$~AU, PA=$-72$\degr, and a mass of $\sim100$~\mo. About 4~arcsec to the south there is a fainter and unresolved source, MM2, of about $\sim50$~\mo.

\item[-] Both MM1 and MM2 are driving high-velocity \co\ outflows, with outflow parameters typical of massive YSOs, and the outflow driven by MM1 shows a  high collimation factor. In addition, both MM1 and MM2 are associated with \tco\ and \ceo\ emission, whose structure, especially for \ceo, is elongated perpendicularly to the outflow directions, similar to toroids seen in other massive star-forming regions.

\item[-] The CS line profile indicates that the object is embedded in dense gas with important contributions of non-thermal motions.


\item[-] We fitted the SED of MM1 using the SED fitting tool of Robitaille \et\ (2007). The best fit yields a foreground extinction of 4.4~mag, an envelope mass of 140~\mo,
and a bolometric luminosity of $\sim10^5$~\lo. For this bolometric luminosity one would expect that the star should emit an ionizing flux high enough to detect the free-free emission at centimetre wavelengths. However, no source associated with IRAS\,19520+2759 is detected at 6~cm, with a rms noise of 0.1~mJy, which could be due to on-going accretion onto the star.

\end{itemize}

Thus, we present evidence that IRAS\,19520+2759 is a luminous YSO embedded in a massive and dense envelope, which is driving a collimated outflow and does not show strong centimetre emission. We propose that this could be an example of a `swollen' or  `bloated' star as proposed by Hosokawa, Yorke, \& Omukai (2010), where accretion produces an increase of the stellar radius, and a decrease of effective temperature and ionizing flux.  Overall, IRAS\,19520+2759 seems to be an excellent accreting massive YSO candidate which could form an O-type star in the future.

\section*{Acknowledgments}

The authors are grateful to the anonymous referee for valuable comments significantly improving the quality of the paper.
AP is grateful to Tom Landecker and Roland Kothes for kindly providing the 21~cm CGPS images, and to Josep Miquel Girart for insightful comments on the paper; RS acknowledges useful discussions with Harold Yorke and Takashi Hosokawa. The Madrid DSCC observations have been done under the Host Country program; the authors acknowledge the kind support of the Robledo staff during such observations.
AP is supported by the Spanish MICINN grant AYA2008-06189-C03 (co-funded with FEDER funds) and by a JAE-Doc CSIC fellowship co-funded with the European Social Fund.
This work has been partially performed at the  Astrophysics Department of the Astrobiology Center (CAB, CSIC/INTA)  and the California Institute of Technology and has been partially  supported by the Spanish MICINN through grants AYA2009-07304 and CONSOLIDER INGENIO 2010 for the team `Molecular Astrophysics: The  Herschel and Alma Era -- ASTROMOL' (ref.: CSD2009-00038). Ongoing development and operations for OVRO and CARMA are supported by the National Science Foundation under a cooperative agreement (grant AST 08-38260). 
RS thanks NASA for partially funding this work by NASA LTSA and ADP  awards (nos. NMO710651/ 399-20-40-06 \& 399-20-40-08); RS also 
received partial support for this work from HST/GO awards  (nos. GO-09463.01, 09801.01, and 10185.01) from the Space Telescope 
Science Institute (operated by the Association of Universities for  Research in Astronomy, under NASA contract NAS5-26555).
This research has made use of the SIMBAD database, operated at CDS, Strasbourg,  France, the NASA's Astrophysics Data System, Aladin, AKARI observations, a JAXA project with the participation of ESA; the Wide-field Infrared Survey Explorer, which is a joint project of the University of California, Los Angeles, and the Jet Propulsion Laboratory/California Institute of Technology, funded by NASA; the Two Micron All Sky Survey, which is a joint project of the University of Massachusetts and the Infrared Processing and Analysis Center/California Institute of Technology, funded by NASA and the National Science Foundation (NSF); the Midcourse Space Experiment, for which processing of the data was funded by the Ballistic Missile Defense Organization with additional support from NASA Office of Space Science; the NASA/ IPAC Infrared Science Archive, which is operated by the Jet Propulsion Laboratory, California Institute of Technology, under contract with NASA; and the Bolocam Galactic Plane Survey, made using Bolocam on the Caltech Submillimeter Observatory, operated by Caltech under a contract from the NSF. Support for the development of Bolocam was provided by NSF grants AST-9980846 and AST-0206158.

{}


\end{document}